# DESIGN OF WIRELESS ELECTRONIC STETHOSCOPE BASED ON ZIGBEE

Ms. Kadam Patil D. D.  and  Mr. Shastri  R. K.

Department of E&TC
Vidya Pratishthan's CoE, Baramati,
Maharashtra, India
kadam.deepali2008@gmail.com , rajveer_shastri@yahoo.com

## ABSTRACT

*Heart sound stethoscope is primary stage to access diseases. In this paper design of an electronic stethoscope with the functions of wireless transmission is discussed. This electronic stethoscope based on embedded processor. The data can be transmitted through wireless transmission using Zigbee module. A microphone is used to pick up the sound of the heart beat. Acoustic stethoscope can be changed into a digital stethoscope by inserting an electric capacity microphone into its head. The signal is processed and amplified to play with or without earphone. Heart sounds are processed, sampled and sent wirelessly using Zigbee module so that multiple doctors can do auscultation. PC connectivity is provided through serial port where from audio and video can be made available through LAN and internet for telemedicine consultation. Heart beat signals are sensed, sent, displayed, monitored, stored, reviewed, and analysed with ease.*

## 1. INTRODUCTION

The Stethoscope is an acoustic medical device for listening to internal sounds in human body which is known, in medical terms, as auscultation. Heart sound auscultation is one of the most basic ways to assess the state of the cardiac function [1]. Some researches concluded that an abnormal heart-rate profile during exercise and recovery is a predictor of sudden death. Because the incidence of cardiovascular disease increased year by year, cardiovascular diseases relating to heart has become worldwide common and high prevalent disease. As a result of the development of wireless technology, the diagnosis based on the analysis of heart sound will become a new method to diagnose cardiovascular disease. Anh Dinh & Tao Wang had processed heart beat signal and sent wirelessly using Zigbee protocol [2]. Some electronic stethoscopes are designed which are using Bluetooth for wireless transmission. At receiver side heart signal can play on earphone and it can be store on PC or PDA [3-6].

One problem with acoustic stethoscopes is that the sound level is extremely low and there are some short comings in the heart sound analysis [3]. 1. The mechanism of the heart sound production is still being debated in the clinical diagnosis. 2. Lack of quantitative analysis techniques or a combination of PCG diagnosis. 3. Auscultation is easily affected by the subjectivity of the doctor and measuring environment. 4. A large amount of heart sound components is low-frequency, which is important for the diagnoses but cannot been clearly distinguished by doctors. 5. The current major clinical application of the heart sound auscultation is a mechanical stethoscope whose accuracy is low [7]. This paper presents wireless electronic stethoscope which overcome these drawbacks. Rest of the paper organised as related work done for this idea, different heart sounds and design of whole system including circuit of signal processing system with its simulation and finally features of system with conclusion.

## 2. RELATED WORK

The development of the stethoscope can be traced back to the beginning of the nineteenth century when a French physician by the name of Rene Laennec first invented the stethoscope in 1816. Heart rate monitoring system with wireless transmission using zigbee is described in [2]. The system includes a bandage size heart beat sensing unit, a wireless communication link, and a networkable computer and a data base. [3] and [4], gives idea about an electronic stethoscope based on embedded processor and Bluetooth transmission which fulfil the shortages from auscultation. It consists of portable device to play heart sound after pre-processing and amplification. In addition, data can be transmitted to PC through Bluetooth. Design of digital stethoscope for heart sound is explained in [9]. The objective of it is to develop a Peripheral Interface Controller based digital Stethoscope to capture the heart sound. The proposed designed device consists of hardware stages like front-end pickup circuitry, microcontroller, graphic LCD and a Serial EEPROM. The captured data can be sent to PC for software analysis using LabVEIW. In electronic stethoscope, main part is heart sound detection which can be studied with the help of [11]. It consists of heart sound detection system based on the new XH-6 sensor to collect the slight heart sound signals, to display in real-time. [18], presents a new concept of home diagnosis system, which is based on an electronic stethoscope and intelligent analyzing software. The system consequently builds a database of patients including their normal S1 and S2; besides a series of heart disease murmurs are also stored as patterns. Data transmission over LAN is described in the paper [19] which proposes a design and implementation of a Web-Based remote digital stethoscope that integrates current software, hardware interface devices, PC, and Internet into the remotely operated virtual instrumentation.

There are several commercially available electronic stethoscopes in the market. One of them is the Littmann Electronic Stethoscope Model 3000 manufactured by 3M [7]. Amplification is up to 18 times greater than the best non-electronic stethoscopes. There is one more electronic stethoscope which is commonly used CEI electronic stethoscope model CE-321 manufactured by C.E.I Technologies. Amplification is up to 18 times greater than standard acoustic scope and with built-in, 8 level volume control.

## 3. HEART SOUNDS

Acoustic heart sounds are produced when the heart muscles open valves to let blood flow from chamber to chamber. A normal heart will produce two heart sounds, S1 and S2 as shown in figure 1. S1 symbolizes the start of systole. The sound is created when the mitral and tricuspid valves close after blood has returned from the body and lungs. S1 is primarily composed of energy in the 30Hz - 45 Hz range. S2 symbolizes the end of systole and the beginning of diastole. The sound is created when the aortic and pulmonic valves close as blood exits the heart to the body and lungs which lie with maximum energy in the 50 Hz - 70 Hz range with higher pitch. Typically, heart sounds and murmurs are of relatively low intensity and are band limited to about 100–1000 Hz. Meanwhile, Speech signal is perceptible to the human hearing. Therefore, auscultation with an acoustic stethoscope is quite difficult [9-10].

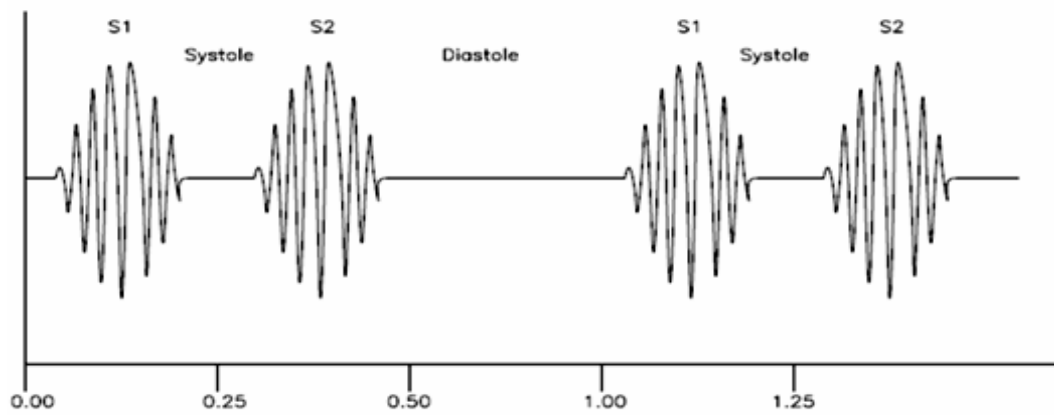

Fig 1: Heart Sounds

## 4. SYSTEM DESIGN

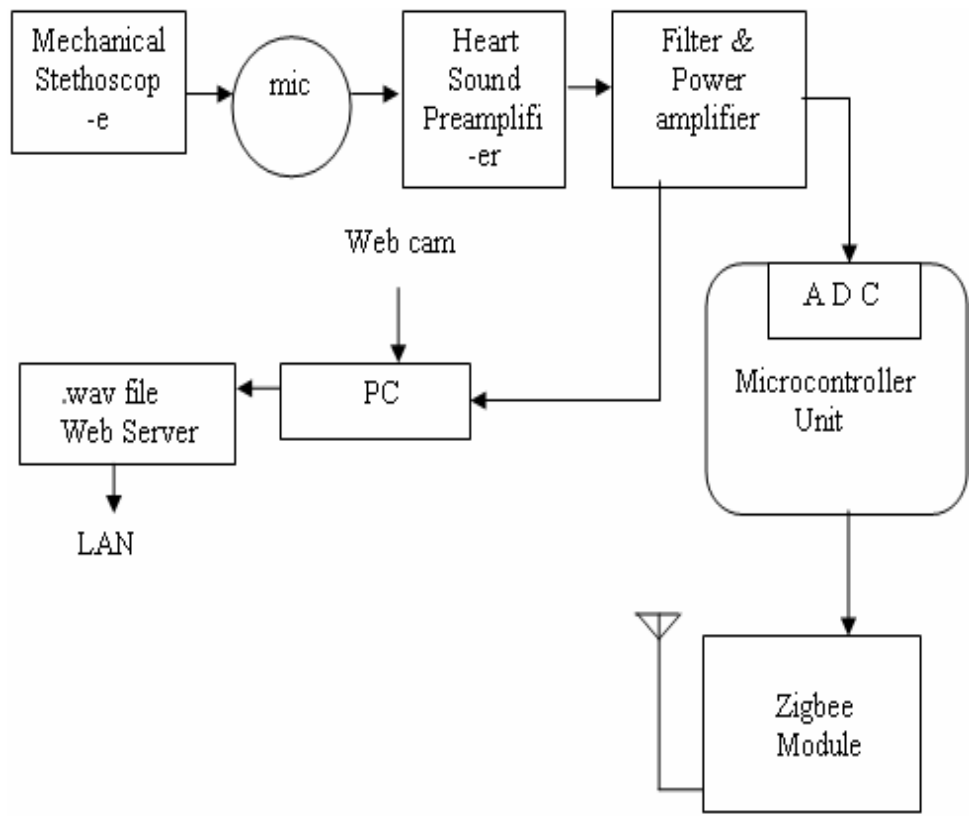

Fig 2: Transmitter

System design consists of two parts that is transmitter and receiver. Fig. 2 shows the transmitter system architecture. The proposed transmitter system consists of the following hardware components: 1) Front end circuitry – sensor, preamplifier, filter and power amplifier with variable gain 2) microcontroller and zigbee module.

## 4.1. Front End Circuitry

Front end circuitry is signal acquisition and preprocessing system [11]. First part is sensor. There are multiple types of sensors that can be used in the chest piece of an electronic stethoscope to convert body sounds into an electronic signal [12]. Microphones and accelerometers are the common choice of sensor for sound recording. Microphone is prefect for the application [13]. The output of the microphone is fed to signal pre-processing module.

Signal pre-processing circuit consists of three parts, which are primary amplification circuit, filter circuit and second amplification circuit [14-15]. The role of signal pre-processing circuit is to adjust the signal from sensor with a series of amplification and filtering so that it meets the follow-up A/D sampling demands and the signal-noise ratio is improved. This circuitry is designed by using operational amplifier [16]. The preamplifier is created to increase the low-signal from the condenser microphone to line-level for further amplification. Here op-amp LM741 is used for designing of preamplifier. It is having gain of 20 which is calculated by feedback resistor value. The output of the preamplifier is fed to an active low pass filter with cut-off of 100 Hz and 1000 Hz so that Heart sounds and respiration sounds are passed and background sounds are reduced. Frequency is selected by selecting capacitor value. Filter is having gain of 1.6. The output signal from the filter is processed by power amplifier to supply the necessary power to drive the headphones for further amplification. The LM386 circuit is an audio amplifier designed for use in low voltage consumer applications which provides both voltage and current gain for signals [17]. Hence power amplifier with variable gain is designed with the help of op-amp LM386. Gain can vary by varying input given to amplifier through pot. Fig 3 shows signal pre-processing circuit.

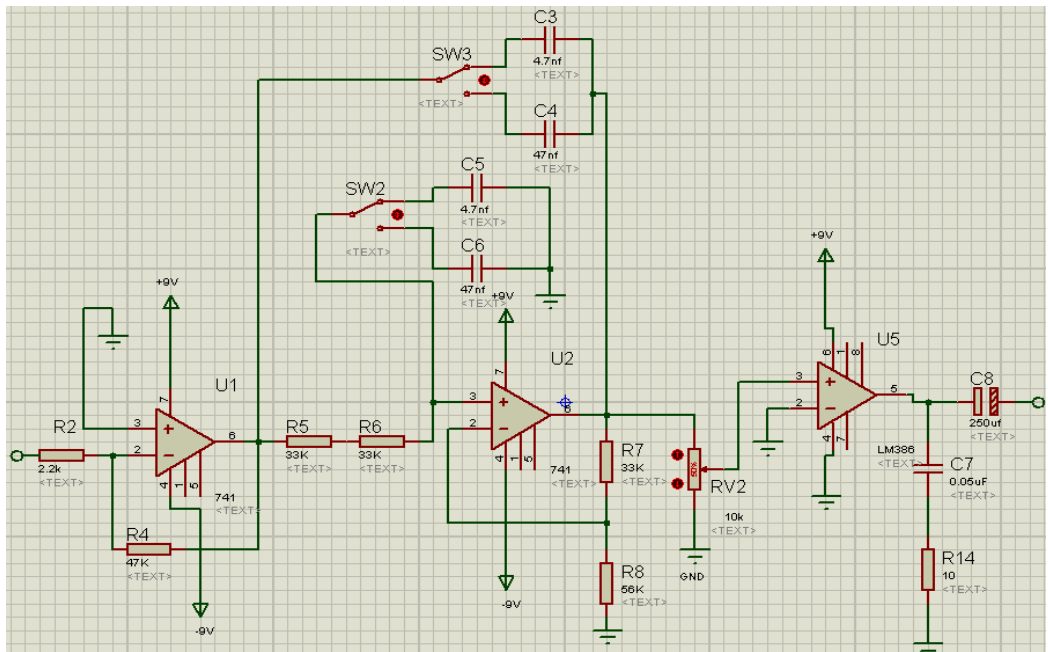

Fig 3: Signal Pre-processing Circuit

## 4.2. Microcontroller and Zigbee module

The output of signal Pre-processing Circuit is converted into digital form by ADC. Inbuilt successive approximation 12 bit ADC of microcontroller is used. Here PIC18f2423 microcontroller is used. Some features are as follows:-

- 0-40 MHz Operating frequency

- 16 Kbytes flash program memory
- 768 bytes data memory
- 12-bit ADC (10 input channels)
- Serial communication :- SSP and USART

For wireless transmission zigbee module JN5148 made by Jennic is preferred. The JN5148-001 is a range of ultra low power, high performance surface mount modules targeted at JenNet and ZigBee PRO networking applications, enabling users to realize products with minimum time to market and at the lowest cost. It's operating frequency is 2.4GHz and data rate is 250 kbps. The modules use Jennic's JN5148 wireless microcontroller to provide a comprehensive solution with large memory, high CPU and radio performance and all RF components included.

### 4.3. PC Connectivity

Signal from conditioner system (analog signal) is given to PC through auxiliary input pin for storage purpose [18-19]. This audio signal is stored in form of **.**wav file for further analysis. This audio video interface is provided using web camera through internet for proper positioning of stethoscope. The LAN support is also provided for this system using JAVA.

### 4.4. Receiver

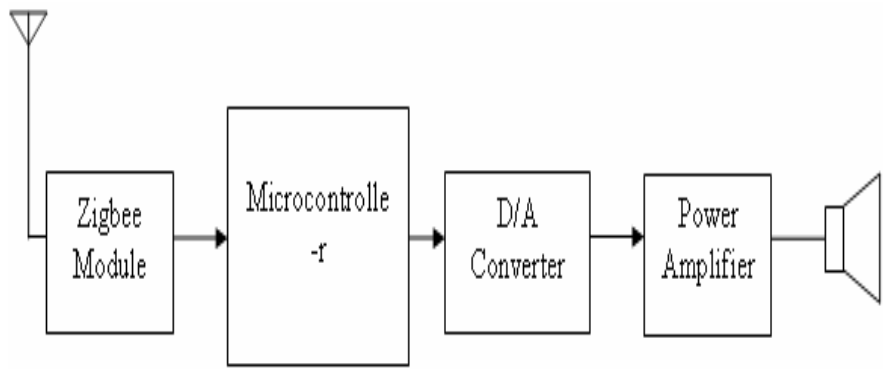

Fig 4:  Receiver

Fig. 4 shows the receiver system. The hardware design of receiver consists of following parts: zigbee module, microcontroller, DAC, Power amplifier. Zigbee module captures the signal in the air and transmits to microcontroller. We have to play this signal on speaker phone. But received signal is in digital form hence we have to first convert it into analog. Hence signal from microcontroller is given to 12 bit digital to analog converter. Here PIC16f873 microcontroller is used. Signal from microcontroller is given to 12 bit DAC MCP4822. The MCP4822 devices are designed to interface directly with the Serial Peripheral Interface (SPI) port available on many microcontrollers. Then this analog signal is amplified by power amplifier with gain control same as at transmitter side. And now this signal is given to speaker. In this way wireless electronic stethoscope system is implemented with provision of heart signal storage on PC for further analysis. This signal is also accessed through over internet for consulting with other physicians. Simulation of signal pre-processing system is done which is discussed in next section.

## 5. SIMUATION OF SIGNAL PRE-PROCESSING SYSTEM

A circuit of signal acquisition and conditioning for electronic stethoscope is designed. With the help of software Proteus **7.6** this circuit has been simulated. Audio file is given as input to circuit and checking for output with the help of oscilloscope. Complete circuit is simulated for heart sound, murmur and different types of lung sounds audio as input for both filter with cut off frequency 100 Hz and 1000 Hz.

**1)** Heart sound audio file is given as input shown in fig 5(a). When it is check for filters with cut off 100 and 1000 Hz, it is noticed that proper amplified output is for 100 Hz frequency filter. Output at different stages is observed. It is shown in fig 5 (b) to fig 5(d).

**2)** Heart sound with late systolic murmurs is given as input as in fig 6 (a) and output is observed by digital oscilloscope at different stages which is shown in fig 6 (b) to fig (d).

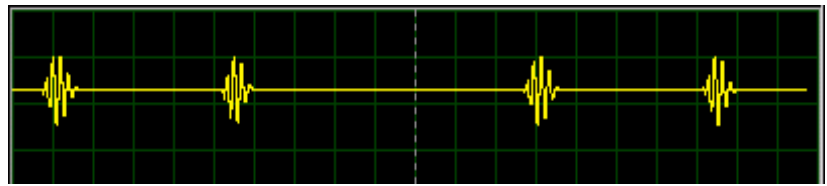

Fig 5 (a): Heart sound as input

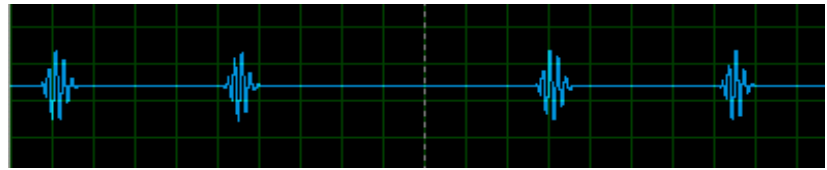

Fig 5 (b): Output at preamplifier stage

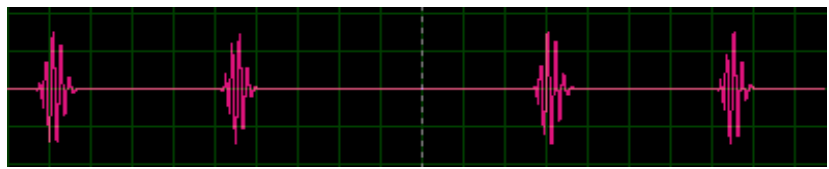

Fig 5 (c): Output at filter stage

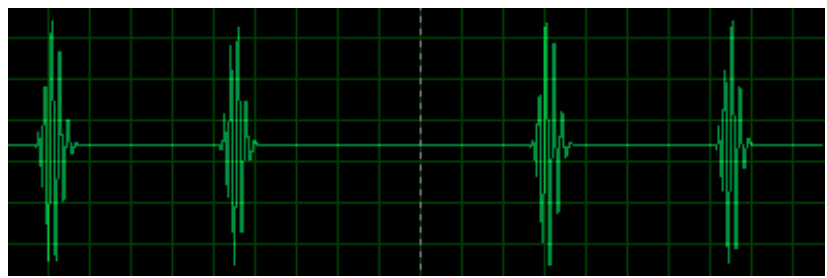

Fig 5 (d): Output at Power amplifier stage

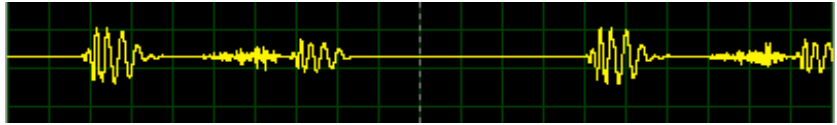

Fig 6 (a): Heart sound with late systolic murmurs as input

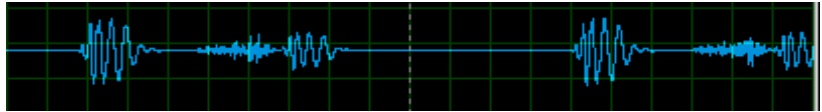

Fig 6 (b): Output for systolic murmurs at preamplifier stage

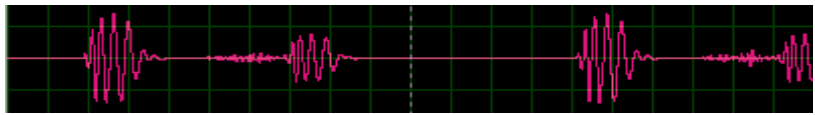

Fig 6 (c): Output for systolic murmurs at filter stage

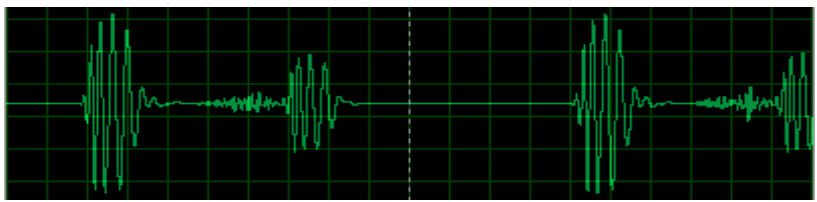

Fig 6 (d): Output for systolic murmurs at Power amplifier stage

**3)** Simulation of circuit is done for different types of lung sounds like normal vesicular lung sound, Inspiratory stridor lung sound, Coarse crackles lung sound, Pleural friction lung sound and Wheezing lung sound. It is observed that there is proper amplified output for filter with cut off 1000 Hz. Results of simulation for normal vesicular lung sound as input are shown. Input is in fig 7 (a). See the changes in output for two filters which are shown in fig 7 (b) and 7 (c). When lung sound is given as input and filter with cut off frequency 1000Hz is selected, better lung sound is obtained than that of when filter with cut off 100 Hz is selected. In same way simulation for other lung sounds (mentioned previous) is done.

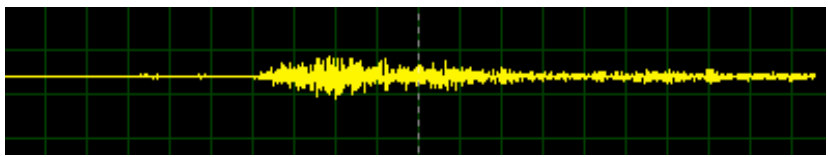

Fig 7 (a): Normal vesicular lung sound as input

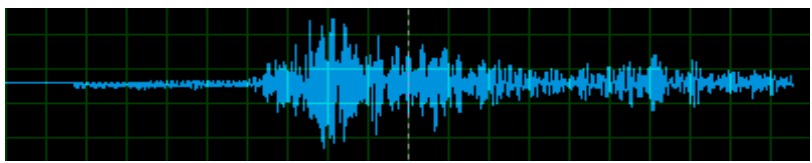

Fig 7 (b): Output at Power amplifier stage when filter with cut off 100Hz

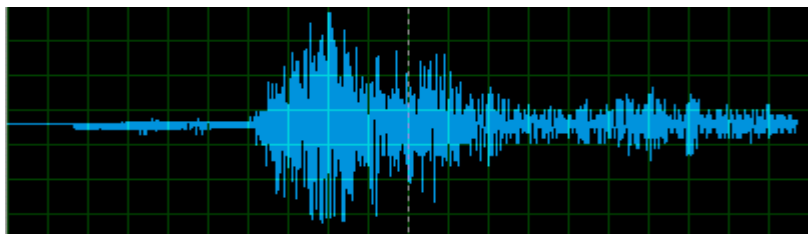

Fig 7 (c): Output at Power amplifier stage when filter with cut off 1000Hz

## 6. FEATURES

Low level heart and lung sounds are amplified with clear audibility so that in noisy area also proper auscultation is possible. Noise reduction takes place by filter that's why accuracy increases. There is gain control facility provided by power amplifier and frequency selection facility provided by filter design. Heart sound can be stored on PC and accessed through internet to consult with other physician. Using Zigbee, wireless auscultation is possible and patient can be monitored by multiple physicians at a time.

## 7. COCLUSION

An embedded digital stethoscope is designed and simulated by using an embedded processor. With the help of PC connectivity, system can also store data and replay for further analysis and consultation. It will help to improve the accuracy of the cardiovascular diseases diagnosis.

Preamplifier is amplifying signal for gain 20. Designed filter is giving proper output until cut off frequency and showing attenuation above that frequency. Frequency selection can be possible by selecting capacitor value with the help of switch. Gain of power amplifier can be controlled by changing value potentiometer connected at input due to which volume control is possible. Signal acquisition and signal pre-processing system of electronic stethoscope which is very important part of system is designed. With the help of Proteus software, circuit of signal pre-processing system is simulated. By simulation results it is clear that the designed circuit gives better heart and lung sounds.

In future, network of multiple transmitters and receivers can be form by using zigbee PRO. When there will be more transmitters, it means diagnosis of heart sound from multiple patients can be possible. As there will be more than one receiver, more than one physician can hear heart sound at a time. It will increase accuracy of diagnosis.

**Authors**

**1) Ms. Kadam Patil D. D. ,** received her B.E degree in Electronics and Telecommunication with distinction in 2008 from Pune University. Currently she is doing M.E. in Electronics (Digital System) from VPCOE, Baramati, Pune University. Her project work included embedded system for application of E-Ticket and Electronic Stethoscope. She had two National Conference Publications.

**2) Mr. Shastri R. K. ,** received the Bachelor of Engineering in 2000 from college of engineering Ambejogai India with distinction, the M.E. Degree(First Class) in Electronics with specialization in computer technology from Shri Guru Govind Singh Engineering and Technology, Nanded India and is now pursuing the Ph.D. degree, in Electronics from Swami Ramanand Tirth Marahawada Univeristy Nanded, India. He has worked as lecturer since 2002 to 2008 and since 2008 he is working as assistant professor in Vidya Pratishthan's college of engineering, Baramati, India. He has published four papers in national conferences and two papers in international journal. He has taught signals and system, digital signal processing, digital image processing, VLSI design and microprocessors. His research interests include biomedical signal processing, and VLSI based image processing.